\documentclass[%
 reprint,
superscriptaddress,
 amsmath,amssymb,
 aps,
prb,
floatfix,
]{revtex4-1}

\usepackage{graphicx}
\usepackage{dcolumn}
\usepackage{bm}
\usepackage[colorlinks=true,allcolors=blue]{hyperref}
\hypersetup{breaklinks=true}

\usepackage{array,booktabs,tabularx}
\usepackage[caption=false]{subfig}
\usepackage[USenglish]{babel}
\usepackage{braket}

\newcommand{\beq}{\begin{equation}}
\newcommand{\eeq}{\end{equation}}
\newcommand{\beqn}{\begin{eqnarray}}
\newcommand{\eeqn}{\end{eqnarray}}

\usepackage[normalem]{ulem}

\begin{document}
\selectlanguage{USenglish}
\preprint{APS/123-QED}

\title{Strong electron-phonon and band structure effects in the optical properties of high pressure metallic hydrogen}

\author{Miguel Borinaga}
\affiliation{Centro de F\'isica de Materiales CFM, CSIC-UPV/EHU, Paseo Manuel de
             Lardizabal 5, 20018 Donostia/San Sebasti\'an, Basque Country, Spain}
\affiliation{Donostia International Physics Center
             (DIPC), Manuel Lardizabal pasealekua 4, 20018 Donostia/San
             Sebasti\'an, Basque Country, Spain}
\author{Julen Iba\~nez-Azpiroz}
\affiliation{Peter Gr{\"u}nberg Institute and Institute for Advanced Simulation, 
             Forschungszentrum J{\"u}lich \& JARA, D-52425 J{\"u}lich, Germany}
\author{Aitor Bergara}
\affiliation{Centro de F\'isica de Materiales CFM, CSIC-UPV/EHU, Paseo Manuel de
             Lardizabal 5, 20018 Donostia/San Sebasti\'an, Basque Country, Spain}
\affiliation{Donostia International Physics Center
             (DIPC), Manuel Lardizabal pasealekua 4, 20018 Donostia/San
             Sebasti\'an, Basque Country, Spain}
\affiliation{Departamento de F\'isica de la Materia Condensada,  University of the Basque Country (UPV/EHU), 48080 Bilbao, 
             Basque Country, Spain}
\author{Ion Errea}
\affiliation{Donostia International Physics Center
             (DIPC), Manuel Lardizabal pasealekua 4, 20018 Donostia/San
             Sebasti\'an, Basque Country, Spain}
\affiliation{Fisika Aplikatua 1 Saila, Bilboko Ingeniaritza Eskola,
             University of the Basque Country (UPV/EHU), Rafael Moreno ``Pitxitxi'' Pasealekua 3, 48013 Bilbao,
             Basque Country, Spain}



\date{\today}

\begin{abstract}
Characterization of the first ever laboratory produced metallic hydrogen 
sample relies on measurements of optical spectra. Here we present 
first-principles 
calculations of the reflectivity 
of hydrogen between 400 and 600 GPa in the $\mathrm{I4_1/amd}$ crystal structure,
the one predicted at these pressures,
based on both time-dependent density functional and Eliashberg 
theories, thus, covering the optical properties from the infrared to the ultraviolet regimes. Our results show that atomic hydrogen displays an interband plasmon at
around 6 eV that abruptly suppresses the reflectivity, 
while the large superconducting gap energy 
yields a sharp decrease of the reflectivity in the infrared region approximately 
at 120 meV. The experimentally estimated electronic scattering rates in the 0.7-3 eV
range are in agreement with our theoretical estimations, which show that the huge 
electron-phonon interaction of the system dominates the electronic scattering
in this energy range.  
The remarkable features in the optical spectra predicted here
encourage to extend the existing 
optical measurements to the infrared and ultraviolet regions.     
\end{abstract}

\maketitle


Wigner and Huntington predicted back in 1935 that at high pressure hydrogen
molecules would dissociate yielding a metallic compound similar to the alkalis~\cite{wigner}.
Metallic hydrogen is expected to be a wonder material as it may superconduct
at ambient temperature~\cite{PhysRevLett.21.1748,cudazzo:257001,PhysRevB.84.144515,Yan20111264,
Maksimov2001569,PhysRevB.93.174308,0953-8984-28-49-494001} and provide 
a very powerful rocket fuel~\cite{1742-6596-215-1-012194}.
Despite a huge experimental effort in the last years has characterized
the phase diagram of hydrogen up to very high pressures~\cite{Eremets2011,Nat.383.702,
PhysRevLett.100.155701,natureliquidh,RevModPhys.66.671,PhysRevLett.108.146402,Goncharov04122001,
Eremets2011,PhysRevLett.108.125501,PhysRevLett.110.217402,Howie2015,optical,Dalladay-Simpson2016},
metallic hydrogen has remained elusive. Remarkably, however,
the long standing quest 
might have come to an end as, early this year, 
Dias and Silvera reported the first 
ever laboratory-produced sample of metallic 
hydrogen~\cite{Diaseaal1579}. 

Metallic hydrogen was claimed to have been observed as the sample
became reflective above 495 GPa~\cite{Diaseaal1579}.
The claim remains controversial as doubts on the pressure calibration
have been raised and a semiconducting sample may also be reflective~\cite{Eremets2017,Loubeyre2017,Liu2017,Goncharov2017}.
Moreover, raw reflectance data in Ref. [\onlinecite{Diaseaal1579}] 
shows a sharp decrease for photon energies larger than 2 eV whose origin is not 
totally understood even though it was first attributed to absorption of diamond\cite{Dias2017}. 
In any case, the claim of 
having produced metallic hydrogen comes after previous works in which the first
signals of its existence were present or close to 
appear~\cite{Dalladay-Simpson2016,Eremets2016}. 
Thus, reproducibility of the experiment and exhaustive characterization of the system 
clearly are the next challenge.
Characterizing hydrogen under pressure is extremely difficult due to the 
limitations imposed on conventional techniques. Unavailability of neutron scattering 
and x-ray diffraction experiments for extremely high-pressure hydrogen samples in Diamond-Anvil-Cells
makes the use of alternative techniques imperative. Many of the already known solid hydrogen 
phases have been characterized by comparing Raman scattering and infrared (IR)
absorption data to theoretical 
calculations\cite{PhysRevLett.108.146402,Goncharov04122001,Eremets2011,
PhysRevLett.108.125501,PhysRevLett.110.217402,Dalladay-Simpson2016,pickard:473}. 
Comparing optical reflectance spectra to theoretical estimations 
is indeed another option~\cite{Diaseaal1579,Perucchi2009}.  

\begin{figure*}[t]
\subfloat[a][]{\includegraphics[height=0.33\linewidth]{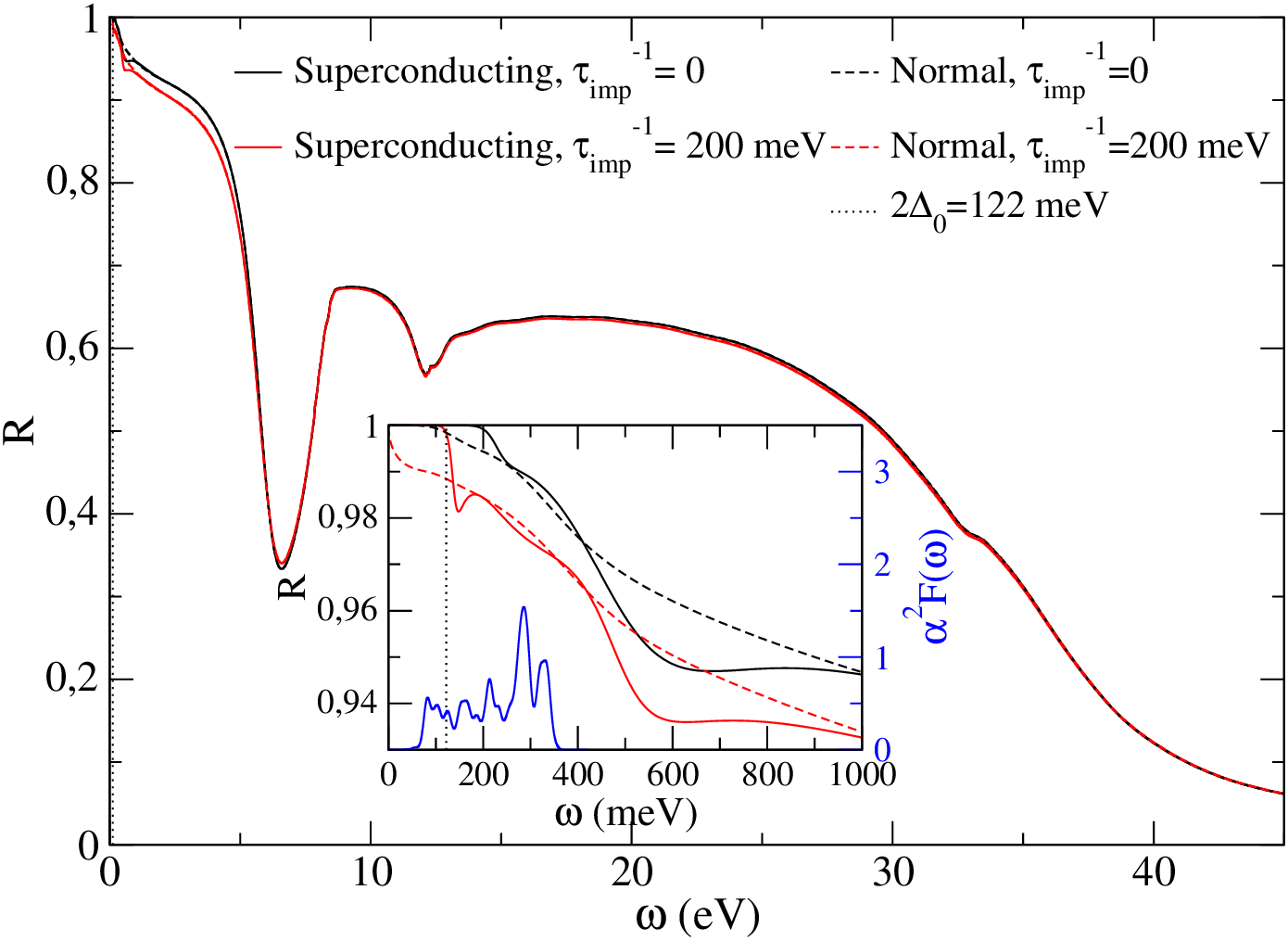}\label{reflectivity}}\hfill
\subfloat[b][]{\includegraphics[height=0.33\linewidth]{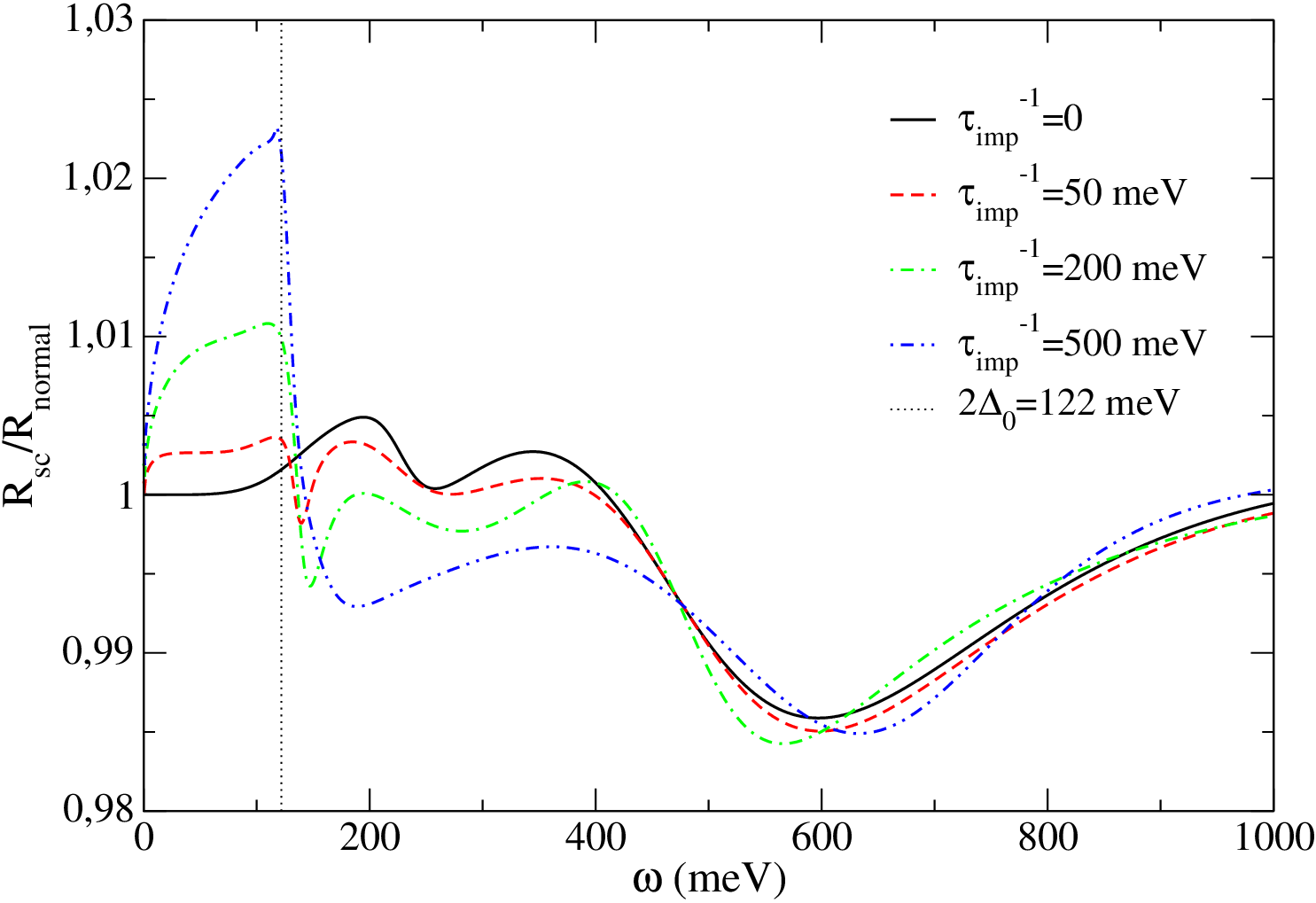}\label{ratio}}
\caption{
(a) Reflectivity of $\mathrm{I4_1/amd}$ hydrogen in vacuum at 50 K and 500 GPa for different 
impurity scattering rates both in the normal and superconducting states. 
The inset shows the same curves at the low energy regime along with the electron-phonon 
spectral function $\alpha^2F(\omega)$. 
(b) Ratio between superconducting and normal state reflectance of 
$\mathrm{I4_1/amd}$ hydrogen in vacuum at 50 K and 500 GPa for different impurity 
scattering rates. 
\label{fig1}}
\end{figure*}

In this letter we report an exhaustive characterization of the optical
response properties from the IR to 
the extreme ultraviolet (UV) of metallic hydrogen between 400 and 600 GPa
in the atomic $\mathrm{I4_1/amd}$ phase, the structure predicted for hydrogen at these 
pressures~\cite{PhysRevLett.112.165501,PhysRevLett.106.165302}.
Our fully first-principles analysis based on density-functional
theory (DFT) sheds light into the regime measured by Dias and Silveira~\cite{Diaseaal1579}, 
from 0.75 to 3 eV photon energies,
showing that in this range the electronic relaxation time is dominated 
by the huge electron-phonon interaction.
Besides, our calculations predict a complex reflectance
spectrum not expected \emph{a priori} for a 
free-electron-like alkali metal.
On one hand, our calculations reveal a sharp onset of the optical conductivity in the IR region
induced by the very large superconducting gap
of atomic hydrogen~\cite{PhysRevB.93.174308}.
This suggests that reflectivity measurements in the IR region
at temperatures below the superconducting
critical temperature $T_c$, predicted to be of 300 K~\cite{PhysRevB.93.174308},
might be used to measure optically $T_c$ and the 
superconducting gap as it occurs, for instance, in 
cuprates~\cite{PhysRevLett.59.1958}, alkali-metal-doped 
fullerenes~\cite{Rotter1992}, and the recently discovered~\cite{Drozdov2015} 
record superconductor H$_3$S~\cite{Capitani2017}. 
On the other extreme, the UV regime 
exhibits a pronounced loss of reflectance due to 
the presence of a non-free-electron-like interband plasmon. 
Our results therefore provide clear means of characterizing metallic hydrogen 
via these singular features, strongly encouraging the extension of experimental 
optical measurements~\cite{Diaseaal1579} to broader regimes.

The central quantity addressed in this work is the frequency dependent reflectivity,
which for normal incident light in a medium with refractive index $n$ can be written as 
\begin{equation}\label{reflec}
 R(\omega)=\left|\frac{\sqrt{\varepsilon(\omega)}-n}{\sqrt{\varepsilon(\omega)}+n}\right|^2.
\end{equation}
We have calculated the relative dielectric function $\varepsilon(\omega)$ by 
combining time-dependent DFT~\cite{PhysRevLett.52.997,PhysRevLett.76.1212,PhysRevB.85.054305} (TDDFT),
which realistically incorporates the actual electronic structure
into the dielectric function, and isotropic Migdal-Eliashberg (ME) equations, 
which take into account  how
an excited electron can decay due to the electron-phonon interaction
(see the Supplementary Material for a more detailed description of the methods and the
calculation procedure). ME equations are solved with different $\tau_{imp}^{-1}$
impurity scattering rates. 
This enables us to properly account for the optical features of metallic 
and presumably superconducting hydrogen not only in the visible and UV, but also in the IR,
which could be strongly affected due to the presence of the 
superconducting gap $\Delta_0$~\cite{PhysRevLett.59.1958,Rotter1992,Capitani2017}. 
All the calculations presented in this letter are performed at 500 GPa, where metallic hydrogen 
is predicted to adopt the $\mathrm{I4_1/amd}$ 
crystal structure~\cite{PhysRevLett.112.165501,PhysRevLett.106.165302}. 
Calculations performed at 400 and 600 GPa presented in the Supplementary Material show 
a very weak pressure dependence of the reflectivity as only minor quantitative  
but not qualitative differences are observed. Thus, the analysis presented here holds
at higher and lower pressures. 

\begin{figure*}[t]
\subfloat[a][]{\includegraphics[width=0.49\linewidth]{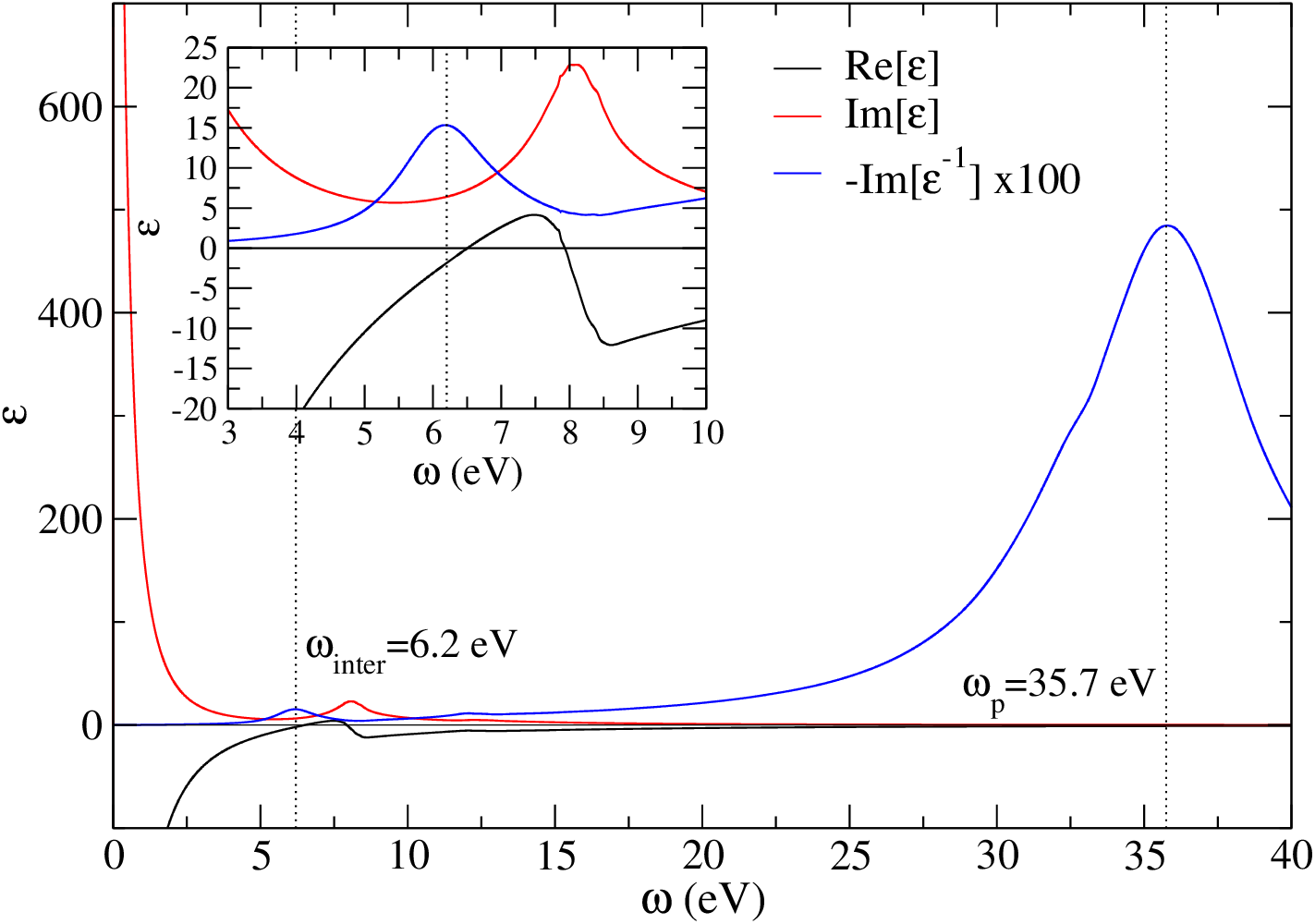}\label{imepsilon}}\hfill
\subfloat[b][]{\includegraphics[width=0.49\linewidth]{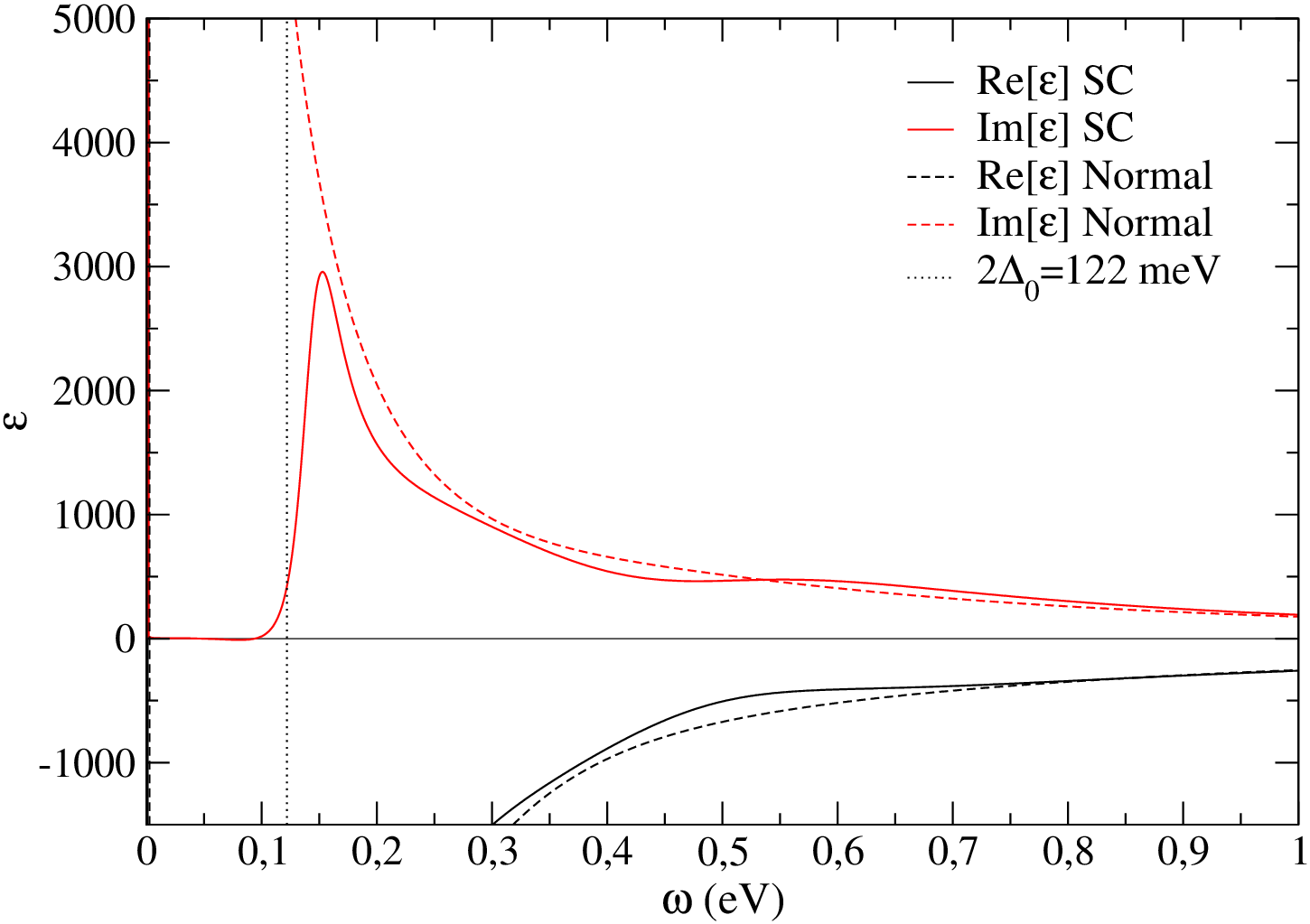}\label{epsilon}}
\caption{(a) Real and imaginary parts of the dielectric function and -Im $\varepsilon^{-1}$ for $\mathrm{I4_1/amd}$ 
hydrogen at 50 K and 500 GPa 
for $\tau_{imp}^{-1}=$ 200 meV in the normal state. 
The inset shows the same curves zoomed in the interband plasmon region.
(b) Real and imaginary parts of the dielectric function of $\mathrm{I4_1/amd}$ 
hydrogen at 50 K and 500 GPa in the IR region
for $\tau_{imp}^{-1}=$ 200 meV in both the normal and superconducting (SC) states. \label{fig2}}
\end{figure*}



In Fig. \ref{reflectivity} we show the calculated reflectivity of $\mathrm{I4_1/amd}$ 
hydrogen in vacuum ($n=1$ in eq. \eqref{reflec}) in the low temperature limit (50 K) 
at 500 GPa, 
for both the normal and superconducting states. We find two different regimes 
for the optical spectra: the IR regime ($\omega<$ 1 eV), where the effects related to 
scattering with phonons and impurities dominate;
and the visible and UV regime ($\omega>$ 1 eV) where
electronic band structure effects start to play a role. 

The inset in Fig. \ref{reflectivity} shows the low temperature limit of the reflectivity 
at 500 GPa for IR radiation. Clean hydrogen ($\tau_{imp}^{-1}=$ 0)  in the normal 
state (which can be obtained by setting $\Delta_0=0$ in ME equations) reflects 
all the incoming light until phonons start
contributing substantially to $\alpha^2F(\omega)$ above $\sim$ 100 meV. When impurity 
scattering is taken into account reflectivity decreases from 1 right from the 
beginning, reaching a small plateau ($\sim$ 0.99 for $\tau_{imp}^{-1}=$ 200 meV)
until scattering with phonons starts to be relevant. In the superconducting state the reflectivity
is equal to unity below $2\Delta_0=$ 122 meV even when impurity scattering is taken into account, 
as that is the amount of energy required to break a 
Cooper pair and make electrons contribute
to the optical conductivity. This can be clearly seen in 
Fig. \ref{epsilon}, where $\text{Im}~\varepsilon$ is strictly zero
below $2\Delta_0$ (except the zero frequency contribution coming from the 
DC conductivity of the Cooper pairs) abruptly
increasing at larger energies. While for $\tau_{imp}^{-1}=$ 200 meV the gap is 
clearly observable due to the sudden decrease of $R$, it is not the same 
for the clean case; in order to have electrons contributing 
to the optical conductivity one needs both 
to break Cooper pairs and scattering with phonons to conserve both 
energy and momentum. The necessity of impurities for observing the 
superconducting gap optically is already well-known, and becomes more 
evident if one plots the ratio between superconducting
and normal state reflectance ($\mathrm{R_{sc}/R_n}$) for different 
$\tau_{imp}^{-1}$ values (Fig. \ref{ratio}). This figure clearly shows 
the emergence of a sharp decrease at $\hbar\omega=2\Delta_0$ only when
impurity scattering is included. The gap is observable even in the clean 
limit ($\tau_{imp}^{-1}=$ 50 meV $<2\Delta_0$), but the drop 
in $\mathrm{R_{sc}/R_n}$ is more notorious as one approaches 
the dirty limit ($\tau_{imp}^{-1}>>2\Delta_0$).

\begin{figure}[t]
 \includegraphics[width=1.0\linewidth]{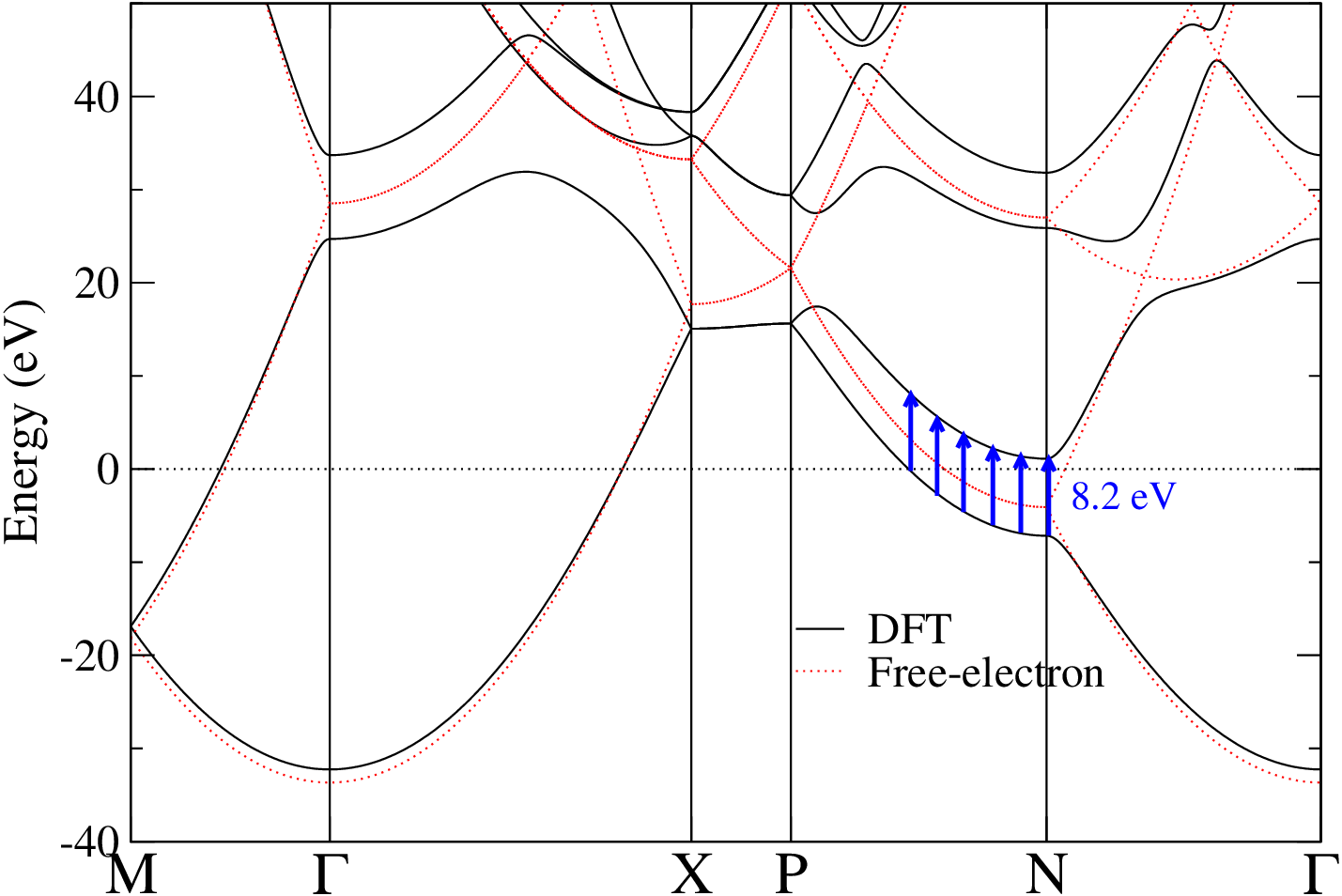}\hspace{0.1cm}
 \caption{\label{bands} Electronic band-structure of $\mathrm{I4_1/amd}$ hydrogen at $500$ GPa.
Interband transitions of about 8.2 eV  around the N point that yield a peak in the imaginary part 
of the dielectric function are marked with blue arrows. The DFT bands are compared to the 
free-electron band structure. The Fermi level is at 0 eV.
}
 \end{figure}

For photon energies larger than 1 eV the reflectivity for normal and superconducting states
are almost identical (see Fig. \ref{reflectivity}). The effect of impurity scattering up to 
5 eV only yields quantitative differences 
keeping the shape of the reflectivity curve
unaltered. In fact, for $\omega>$ 5 eV all the curves both in the normal 
and superconducting state converge into one, suggesting
electronic scattering is dominated by electronic band structure effects 
rather than phonons and impurities.
Remarkably, in this UV regime the reflectivity sharply decreases from a  
high $\sim$ 0.95 value in the visible range ($\hbar\omega=$ [1.6-3.3] eV) 
to $\sim$ 0.3 at $\hbar\omega=$ 6.5 eV. 
This stark reduction of the reflectance is a consequence of light absorption 
due to the presence of an interband plasmon
not expected \emph{a priori} for a simple free-electron-like 
alkali metal. This is demonstrated in Fig. \ref{imepsilon}, where we display the calculated dielectric function. 
As revealed in the inset of this figure, the interband plasmon emerges around the energy where the real part of $\epsilon(\omega)$ vanishes
and its imaginary part remains low. 
This induces a clear peak in $-\text{Im}~\varepsilon^{-1}$ at $\omega_{inter}=6.2$ 
eV as shown in Fig. \ref{imepsilon}, which coincides with
the drastic drop in the reflectivity.
We label this plasmon as interband because it is a consequence
of the interband transitions of around 8.2 eV
that occur close to the N point (see band structure in Fig. \ref{bands}).
Consequently, the imaginary part of the dielectric function shows
a clear peak at 8.2 eV, which due to Kramers-Kronig relations
makes the real part pass through 0 at 6.5 eV and create the
interband plasmon.
Even if the band structure of $\mathrm{I4_1/amd}$
is not far from the free-electron limit, the large gap
opened by the electron-ion interaction at the N point~\cite{PhysRevB.93.174308}
suffices to induce the presence of an interband plasmon not expected
for a free electron-like metal. Thus, metallic hydrogen
in the $\mathrm{I4_1/amd}$ phase is another example in which
the departure from the free-electron-like character
makes interband plasmons emerge and abruptly modify  
the optical properties, as it occurs in other simple 
compounds under pressure such as Li~\cite{silkin:172102,1367-2630-10-5-053035,PhysRevB.81.205105}, 
Ca~\cite{PhysRevB.86.085106}, Na~\cite{PhysRevB.89.085102,Mao20122011}, 
and AlH$_3$~\cite{PhysRevB.82.085113}. 

Apart from the interband plasmon, we find that metallic hydrogen
shows the expected free-electron plasmon at $\omega_p=$ 35.7 eV,
which is responsible for the final decrease of the reflectivity 
at the extreme UV regime. This value is in good agreement 
with the Drude-model estimate of 
$\omega_p = \sqrt{4\pi N}=$ 35.0 eV,
where $N$ is the total electronic density,
and the 33.2 $\pm$ 3.5 eV
experimental value obtained by Dias and Silvera by
fitting to a Drude model only the two lowest-energy points,
which are not affected 
by diamond absorption\cite{Dias2017}. 
This experimental fit also provided a 
$\tau^{-1} = 1.1 \pm 0.2$ eV estimate for the electronic scattering rate\cite{Dias2017}. 
In order to shed some light into these experimental values 
we fit our calculated dielectric function to a Drude model
with frequency-dependent $\omega_p(\omega)$ and $\tau^{-1}(\omega)$, which are displayed in Fig. \ref{tau} 
(see Supplementary Material for details). 
In the $\hbar\omega= [0.7-3]$ eV range our results
yield $\omega_p\sim$ 21 eV and $\tau^{-1}\sim 0.6-1$ eV at 5 K for 
a clean sample, with impurities shifting $\tau^{-1}$ upwards.
Our estimated $\tau^{-1}$ in the measured frequency range is in good agreement
with the experiment and clearly shows that its large value  
is mainly due to the strong electron-phonon interaction, which highlights the fact that we are dealing with a superconductor
with a very large $\mathrm{T_c}$.  
Indeed, the electron-electron scattering contribution to $\tau^{-1}$
is negligible in this frequency range as shown by the pure 
TDDFT calculation. 
The value obtained for $\omega_p$ is considerably lower than the 
one obtained by Silvera and Dias\cite{Diaseaal1579,Dias2017}, but it is consistent
with the $\omega_{p}^{intra}=\sqrt{4 \pi N_{intra}}=$ 22.6 eV
value, where $N_{intra}$ is the electronic density contributing to intraband 
transitions (see Supplementary Material for details).
Nonetheless, in this low energy
($\omega<<\omega_p$) regime very different $\omega_p$ values still provide 
a good fitting to the experimental data, while $\tau^{-1}$ remains almost unaltered
(see Supplementary Material). 
We thus consider the $\tau^{-1}$ value 
obtained experimentally\cite{Diaseaal1579,Dias2017} to be more meaningful
than the plasma frequency, because indeed it is this parameter what determines
how much the reflectivity deviates from one for $\omega << \omega_p$. 




\begin{figure*}[t]
\subfloat[a][]{\includegraphics[height=0.34\linewidth]{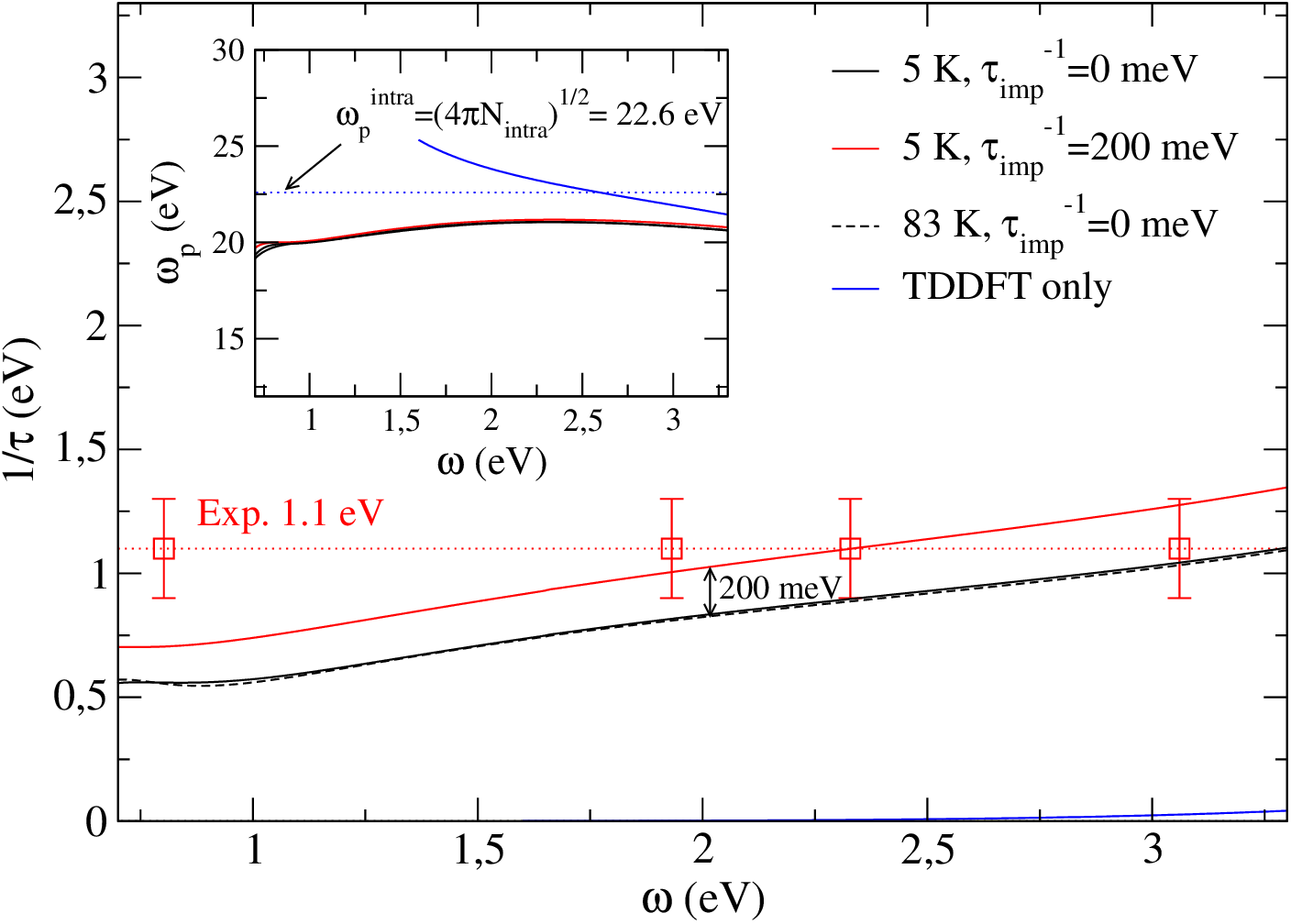}\label{tau}}\hfill
\subfloat[b][]{\includegraphics[height=0.34\linewidth]{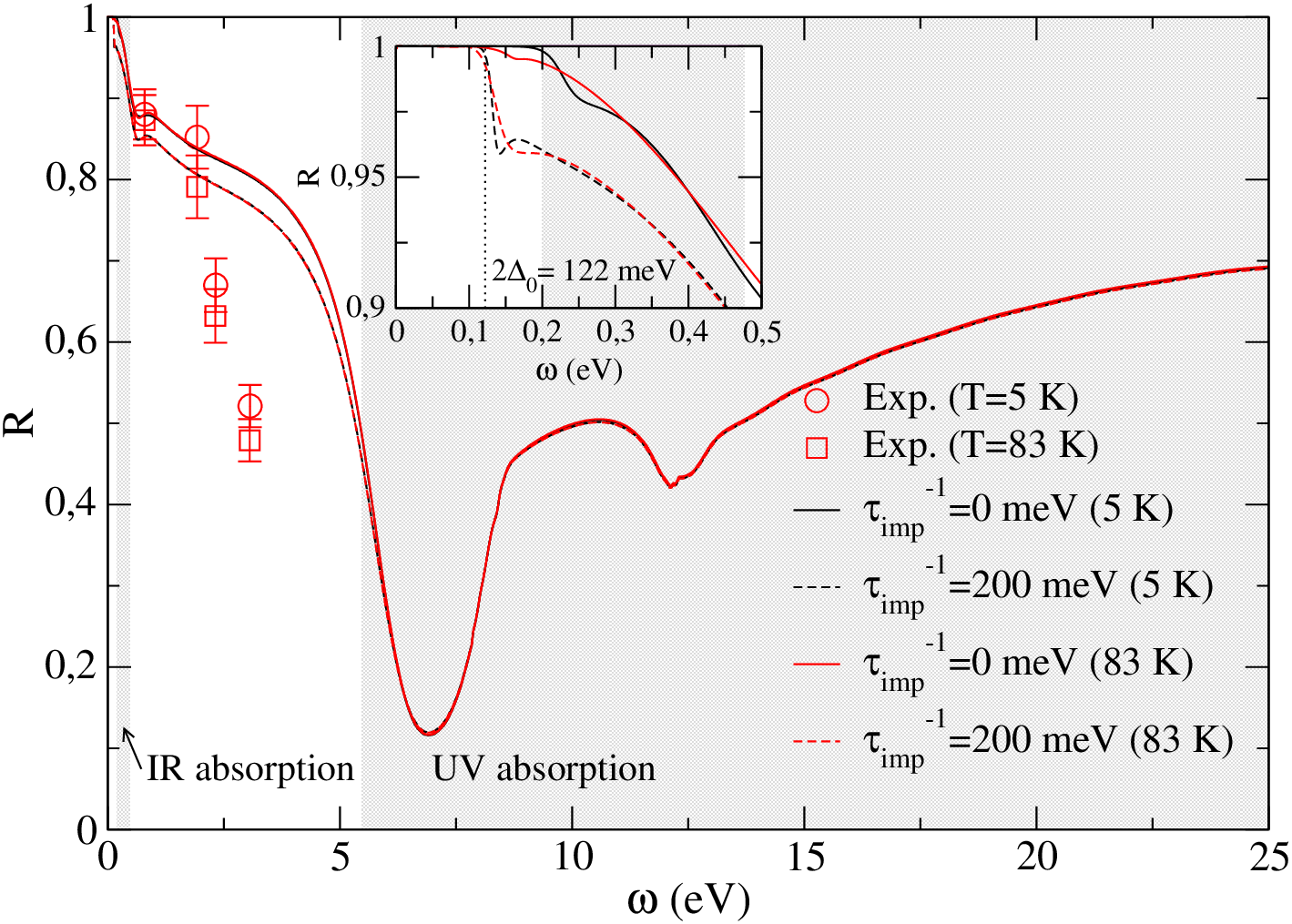}\label{exp}}
\caption{(a) Frequency-dependent electronic scattering rate of $\mathrm{I4_1/amd}$ hydrogen 
at $500$ GPa for different impurity scattering rates and temperatures in the region in which the experiments in Ref. 
[\onlinecite{Diaseaal1579}] were performed. The same curve is also obtained for the case in which the ME formalism is
 not considered. The experimental $\tau^{-1} = 1.1 \pm 0.2$ eV estimate is also included\cite{Dias2017}.
The inset shows the obtained frequency dependent $\omega_p(\omega)$ in the same energy range, 
together with the $\omega_p^{intra}$ estimate. 
(b) Reflectivity of a $\mathrm{I4_1/amd}$ hydrogen/diamond interface at 5 and 83 K at 500 GPa for 
different impurity scattering rates. Diamond IR (0.2-0.47 eV) and UV (5.47 eV electronic bandgap) 
absorption  regions are shown in shaded gray. 
Experimental raw values~\cite{Diaseaal1579} are included.\label{experiments}}
\end{figure*}

In Fig. \ref{exp} we show how the raw experimental values in Ref. [\onlinecite{Diaseaal1579}] compare to our
calculations at 5 and 83 K, with reflectivity calculated for a hydrogen/diamond interface 
by using the refractive index of $n=$2.41 of diamond instead of $n=$1 in eq. \ref{reflec}. 
Our results compare well at 5 K to the two lowest frequency experimental data points, 
which are the ones considered for fitting $\omega_p$ and $\tau^{-1}$ with the 
Drude model as the other points might have been affected by absorption of 
light by diamond\cite{Diaseaal1579,Dias2017}.
The sharp offset of reflectivity due to the superconducting gap lays off the IR absorption range of diamond, 
and should be measurable in consequence.
UV absorption of diamond however would eclipse the minimum of the reflectivity
predicted here at 6.5 eV due to the presence of the 
interband plasmon, since above the indirect electronic bandgap of 5.47 eV (at zero pressure)
diamond is no longer transparent. 
In any case, the sharp decrease associated to such plasmon starts before the absorption onset
and should be observable in pure diamond. Interestingly, recent studies claim the diamond 
bandgap increases with pressure\cite{Gamboa2016}, which would make the 
reflectivity drop induced by the interband plasmon easier to observe. 
However, impurities in diamond could be responsible of light absorption at lower
energies, even in the visible\cite{Dias2017}.
In order to disentangle
whether the reflectivity drop observed experimentally is a consequence
of diamond absorption or reflects the presence of the intraband
plasmon we are predicting here, a
proper characterization of the optical properties of diamond
at the experimental conditions is required.
 
Regarding the temperature dependence of the reflectivity observed in the experiment, 
according to our calculations temperature only affects the region within some meVs 
around the superconducting gap. 
In the experimental region our calculations are practically temperature independent, 
as it can be seen for both reflectivity and scattering rate values (see Fig. \ref{experiments}).
This indicates the temperature dependence of reflectivity shown in 
the experiments cannot be explained with the increase of phonon occupation in the system.  
Motivated by uncertainties in the reported pressure of the 
experiment~\cite{Eremets2017,Loubeyre2017,Liu2017,Goncharov2017}, 
we have also calculated the optical spectra for $\mathrm{I4_1/amd}$ hydrogen 
at 400 and 600 GPa and only found minor quantitative differences with 
respect to the 500 GPa results analyzed in detail here, so that the analysis holds. 
The energy of the interband plamon at 400 GPa is 5.5 eV and
6.6 eV at 600 GPa (more details on the Supplementary Material).


In conclusion, we have made an exhaustive analysis of the optical
response properties of $\mathrm{I4_1/amd}$ metallic hydrogen from the infrared to 
the extreme ultraviolet. Our results show that in the measured
energy range\cite{Diaseaal1579} the electronic scattering is dominated by the huge electron-phonon interaction
of the system. Besides, our calculations reveal a sharp onset of the optical conductivity in the infrared region
induced by the very large superconducting gap and a pronounced loss of reflectance in the ultraviolet regime due to 
the presence of a non-free-electron-like interband plasmon. 
Thus, our work deeply encourages further experimental research in order 
to extend optical measurements both to the ultraviolet and infrared. Confirming
the predicted interband plasmon and measuring the superconducting 
gap optically would be not only of tremendous interest by itself, but 
also a big step towards characterizing this fascinating material.  


\bigskip

The authors acknowledge financial support from the
Spanish Ministry of Economy and Competitiveness (FIS2016-76617-P) 
and the Department of Education, Universities and Research of the Basque 
Government and the University of the Basque Country (IT756-13).
M.B. is also thankful to the Department
of Education, Language Policy and Culture of the Basque
Government for a predoctoral fellowship (Grant No. PRE-2015-2-0269). 
Computer facilities were provided by the 
Donostia International Physics Center (DIPC). JIA acknowledges the Impuls und
Vernetzungsfonds der Helmholtz-Gemeinschaft Postdoc Programme.

\bibliography{bibliografia}
\end{document}


\selectlanguage{USenglish}



\title{ \textbf{Supplementary Material} \\~ \\ Strong electron-phonon and band structure effects in the optical properties of high pressure metallic hydrogen}

\author[1,2]{Miguel Borinaga}
\author[3]{Julen Iba\~nez-Azpiroz}
\author[1,2,4]{Aitor Bergara}
\author[2,5]{Ion Errea}
\affil[1]{Centro de F\'isica de Materiales CFM, CSIC-UPV/EHU, Paseo Manuel de
             Lardizabal 5, 20018 Donostia/San Sebasti\'an, Basque Country, Spain}
\affil[2]{Donostia International Physics Center
             (DIPC), Manuel Lardizabal pasealekua 4, 20018 Donostia/San
             Sebasti\'an, Basque Country, Spain}
\affil[3]{Peter Gr{\"u}nberg Institute and Institute for Advanced Simulation, 
             Forschungszentrum J{\"u}lich \& JARA, D-52425 J{\"u}lich, Germany}             
\affil[4]{Departamento de F\'isica de la Materia Condensada,  University of the Basque Country (UPV/EHU), 48080 Bilbao, 
             Basque Country, Spain}
\affil[5]{Fisika Aplikatua 1 Saila, Bilboko Ingeniaritza Eskola,
             University of the Basque Country (UPV/EHU), Rafael Moreno ``Pitxitxi'' Pasealekua 3, 48013 Bilbao,
             Basque Country, Spain}

\renewcommand\Authands{ and }
\renewcommand*{\Affilfont}{\small\itshape}

\date{\vspace{-5ex}}
\maketitle

\section{Calculation methods and procedure}

\begin{figure}[t]
\subfloat[a][]{\includegraphics[width=0.49\linewidth]{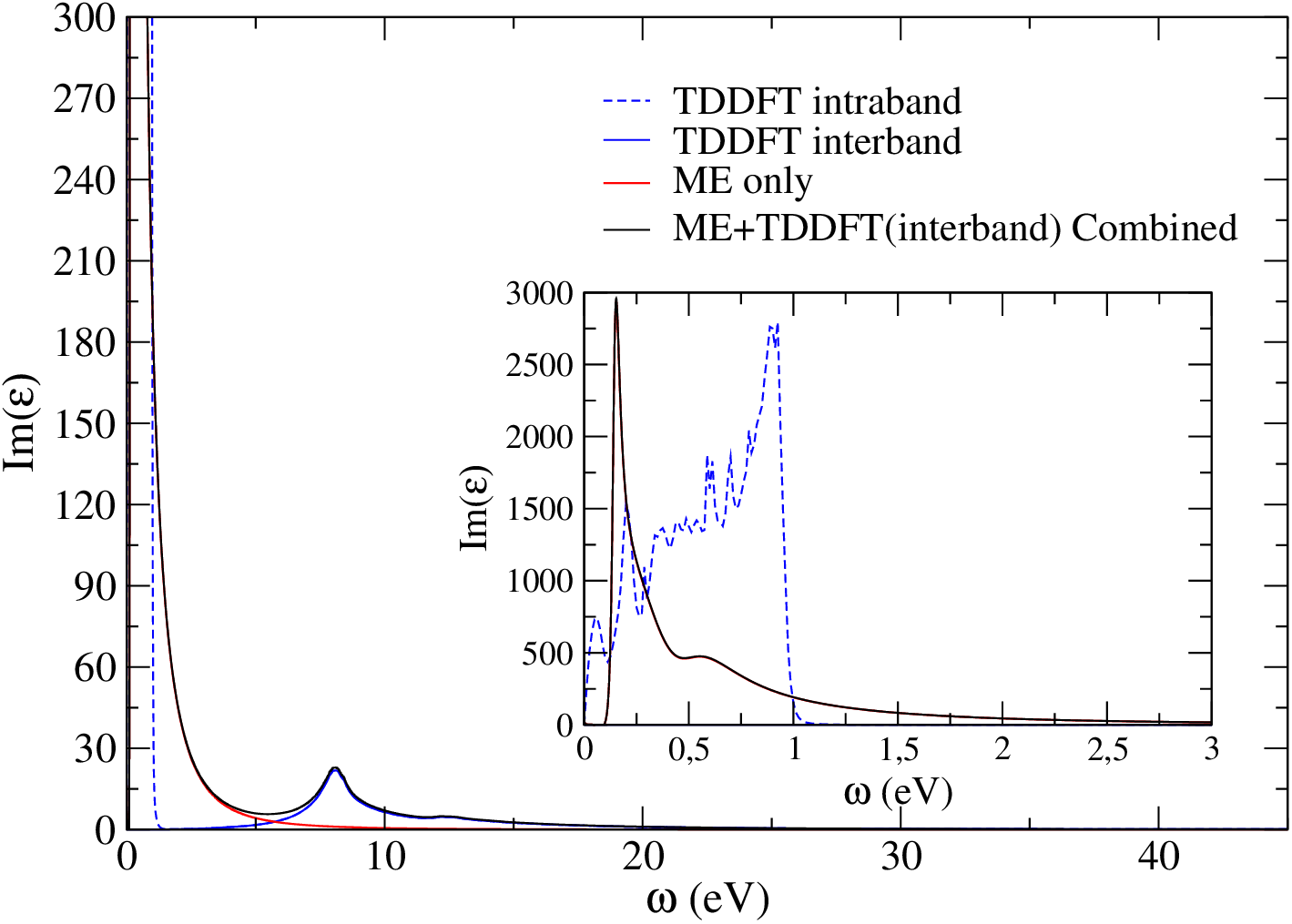}\label{imepsilon}}\hfill
\subfloat[b][]{\includegraphics[width=0.49\linewidth]{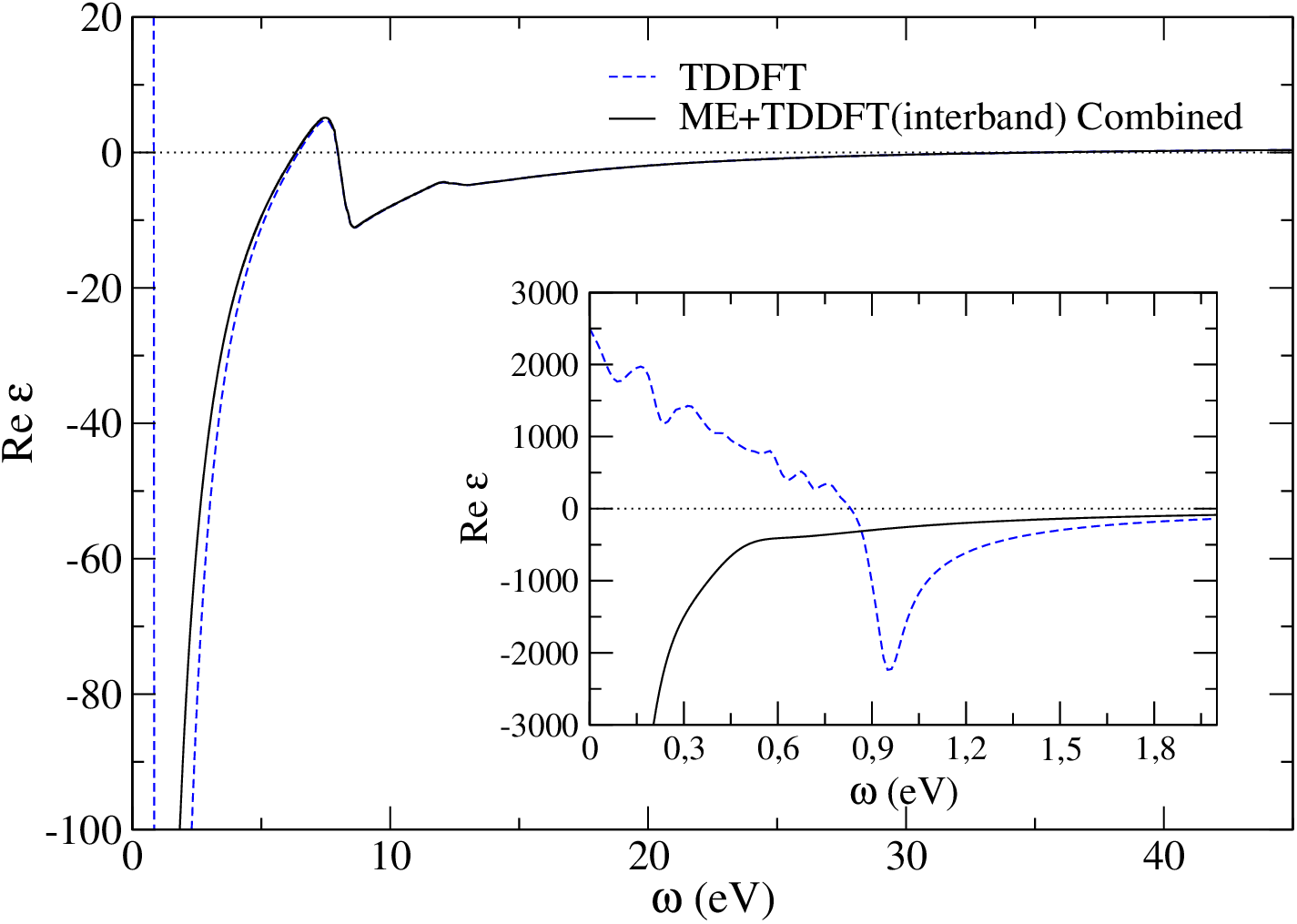}\label{reepsilon}}
\caption{
(a) Different contributions to the imaginary part of $\varepsilon$ of of $\mathrm{I4_1/amd}$ hydrogen at $500$ GPa. The inset shows a zoom into lower energies.
(b) Real part of $\varepsilon$ calculated using the Kramers-Kronig relations.
\label{fig1}}
\end{figure}

The central quantity addressed in this work is the frequency dependent reflectivity,
which for normal incident light in a medium with refractive index $n$ can be written as 
\begin{equation}\label{reflec}
 R(\omega)=\left|\frac{\sqrt{\varepsilon(\omega)}-n}{\sqrt{\varepsilon(\omega)}+n}\right|^2,
\end{equation}
where the relative dielectric function $\varepsilon(\omega)$ can be expressed in terms of the 
optical conductivity $\sigma(\omega)$ as
\begin{equation}\label{eq:dielectric}
 \varepsilon(\omega)=1+i\frac{4\pi\sigma(\omega)}{\omega}.
\end{equation}
The optical conductivity of a metal can be described as 
\begin{equation}
\sigma(\omega)=\sigma_{intra}(\omega)+\sigma_{inter}(\omega)+\sigma_{phonons}(\omega),
\label{opcond}
\end{equation}
where $\sigma_{intra}$ and $\sigma_{inter}$ account, respectively,
for the optical conductivity provided by electronic intraband and interband
transitions, while $\sigma_{phonons}$ accounts for the direct phonon
absorption contribution.
As $\mathrm{I4_1/amd}$ hydrogen lacks of IR active vibrational 
modes~\cite{PhysRevB.93.174308}, we set 
$\sigma_{phonons}=0.$ 

\begin{figure}[t]
 \includegraphics[width=0.7\linewidth]{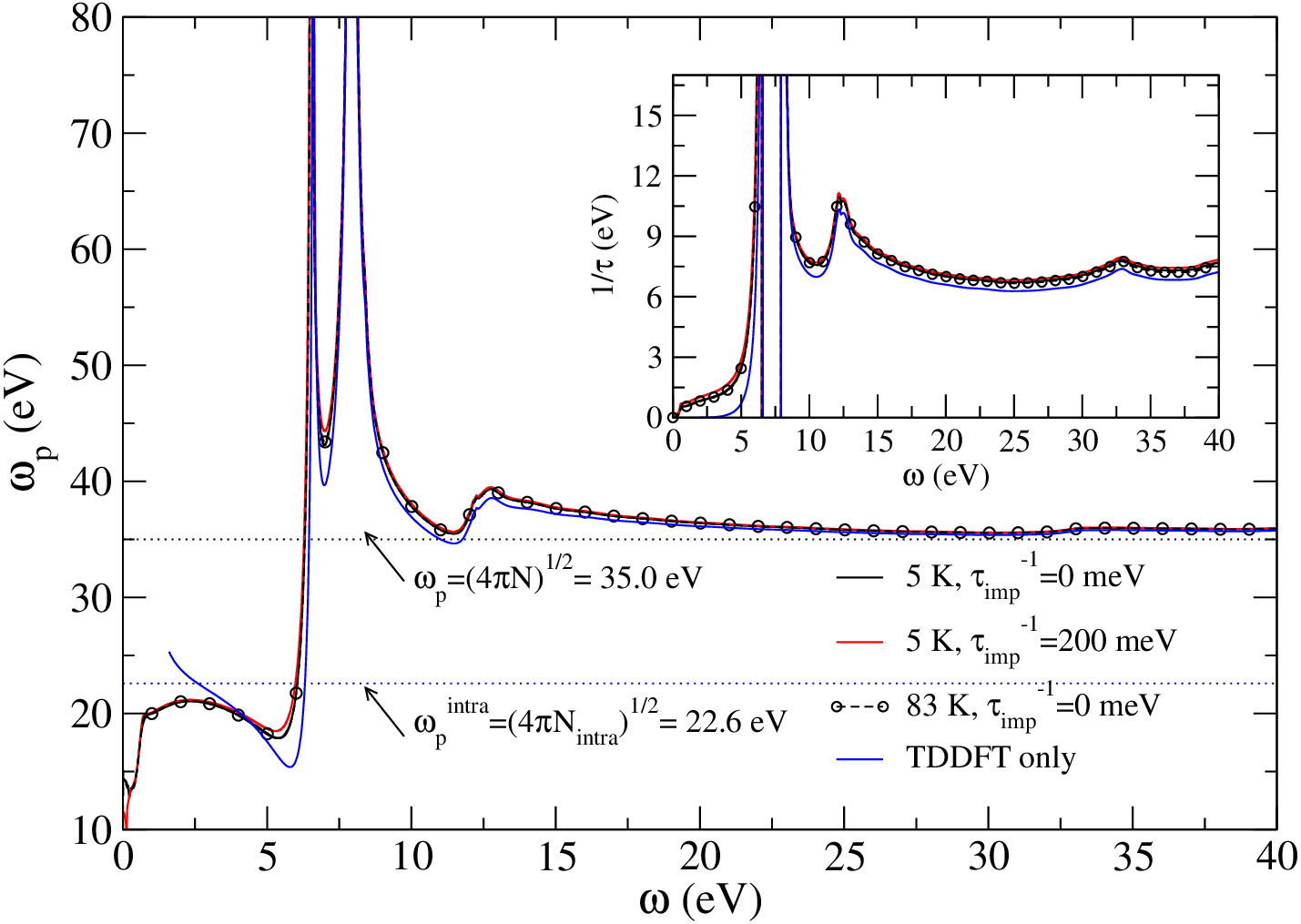}\hspace{0.1cm}
 \caption{\label{tauwide} Drude model frequency-dependent plasma frequency $\omega_p(\omega)$ of $\mathrm{I4_1/amd}$ hydrogen at $500$ GPa for different impurity scattering rates and temperatures. The same curve is also obtained for the case in which the ME formalism is
 not considered (TDDFT only). The inset shows the frequency-dependent impurity scattering rate $\tau^{-1}(\omega)$.
}
\end{figure}

The interband and intraband contributions are computed in two stages. 
We first calculate
the dielectric function within time-dependent DFT~\cite{PhysRevLett.52.997,PhysRevLett.76.1212} (TDDFT),
which realistically incorporates the actual electronic structure
into the dielectric function. 
The dielectric function is calculated by employing  
an interpolation scheme~\cite{PhysRevB.85.054305,PhysRevB.86.085106,PhysRevB.89.085102} of both the Kohn-Sham states
and the matrix elements with the use of maximally localized Wannier functions~\cite{PhysRevB.56.12847,marzari2012}.
The method allows a very fine sampling of the reciprocal space.
In order to avoid numerical problems, a finite but small momentum
is taken for the calculation of the dielectric function. The obtained optical conductivity
from the TDDFT calculation thus contains both interband and intraband
contributions: $\sigma^{TDDFT}(\omega) = \sigma_{intra}^{TDDFT}(\omega)
+ \sigma_{inter}^{TDDFT}(\omega)$.
In order to incorporate
the fine features of the band structure,
we set $\sigma_{inter}(\omega)=\sigma_{inter}^{TDDFT}(\omega)$ in Eq. \eqref{opcond},
which provides a fine description of the reflectivity at high energies.
The low-energy intraband contribution given by
$\sigma_{intra}^{TDDFT}(\omega)$ is affected by the choice
of a finite momentum and completely neglects how
an excited electron can decay 
due to the electron-phonon interaction. Moreover, this regime can also be 
strongly affected in superconductors due to the presence of the 
superconducting gap~\cite{PhysRevLett.59.1958,Rotter1992,Capitani2016}.
In order to incorporate these effects into the reflectivity, the intraband
contribution to the optical conductivity is calculated instead by
solving the isotropic
Migdal-Eliashberg (ME) equations. We thus make
$\sigma_{intra}(\omega)=\sigma^{ME}(\omega)$,
where $\sigma^{ME}(\omega)$ is the optical conductivity obtained
solving ME equations. 
Following Ref.~\cite{PhysRevB.42.67},
these equations are solved 
in the imaginary axis using a Pad\'e approximant for the analytic continuation 
to the real frequency axis.
The free-electron density used in the 
ME equations is determined so that the partial f-sum rule integral
\begin{equation}
 \int_{0}^{\infty}Re~\sigma^{ME}(\omega)d\omega= \int_{0}^{\infty}Re~\sigma_{intra}^{TDDFT}(\omega)d\omega
\label{fsum}
\end{equation}
is satisfied by $\sigma^{ME}(\omega)$. This guarantees that 
the total optical conductivity in Eq. \eqref{opcond} 
satisfies the f-sum rule and yields the correct electron density.
This procedure can be performed because the interband and intraband
contributions in the TDDFT optical conductivities (real part) are
clearly separated as it can be seen in Fig. \ref{imepsilon} for Im $\varepsilon(\omega)=\omega\mathrm{Re~}\sigma(\omega)/4\pi$. The real part of $\epsilon$ (Fig. \ref{reepsilon})
is obtained afterwards using the Kramers-Kronig relations.   
Performing the f-sum rule integral (Eq. \eqref{fsum}) for each contribution in the TDDFT calculation
we obtain $\omega_p^{intra}=\sqrt{4 \pi N_{intra}}=$ 22.6 eV and $\omega_p^{inter}=\sqrt{4 \pi N_{intra}}=$ 26.3 eV, with
$\omega_p^=\sqrt{4 \pi N}=\sqrt{4 \pi (N_{intra}+N_{inter})}=$ 35.0 eV, with $N_{intra}$ and $N_{inter}$ being the electronic density contributing to the intraband
and interband processes respectively.

\begin{figure}[t]
 \includegraphics[width=0.70\linewidth]{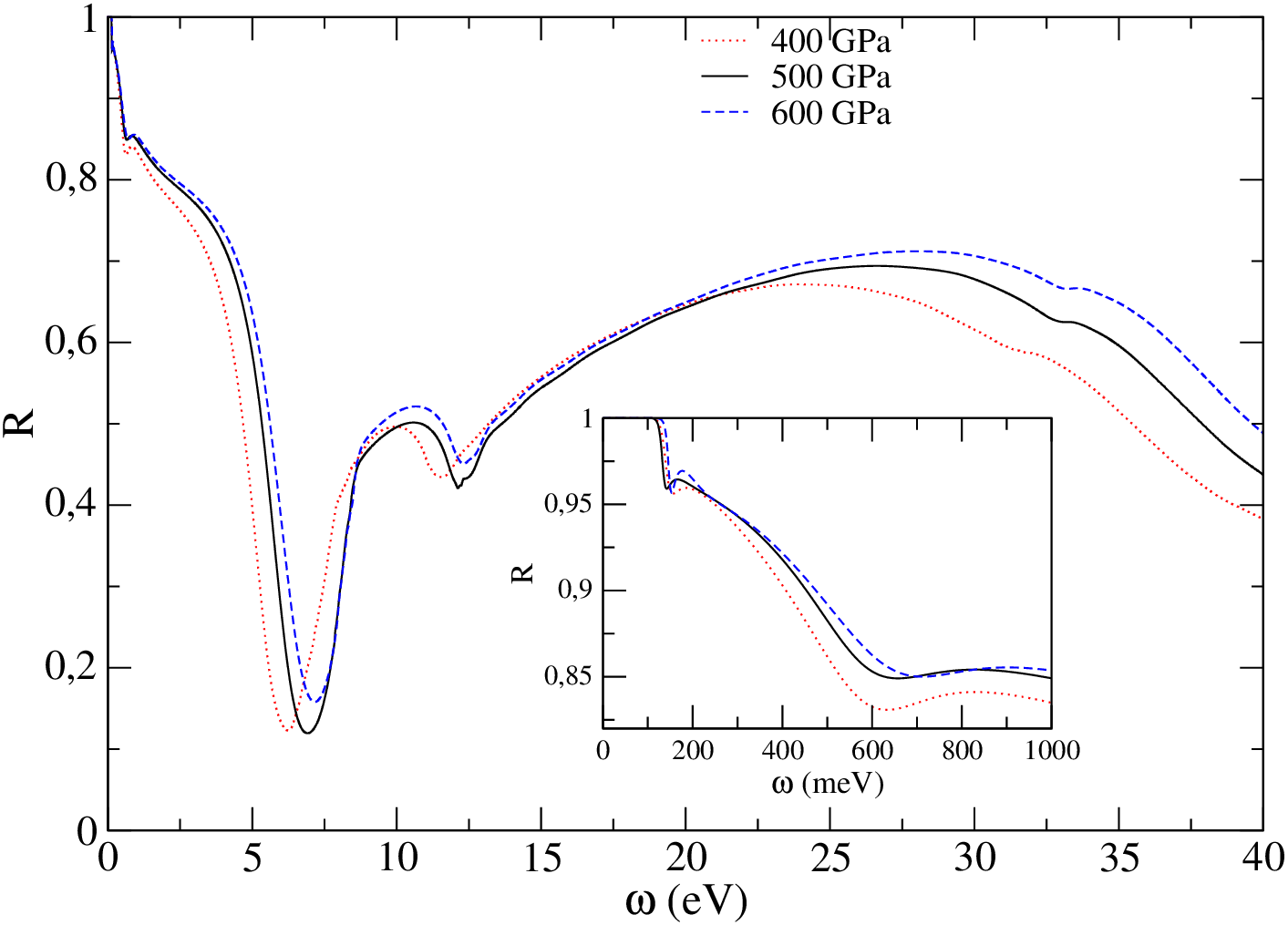}\hspace{0.1cm}
 \caption{\label{pressure} Reflectivity of a $\mathrm{I4_1/amd}$ hydrogen/diamond interface at 400,500 and 600 GPa at 5 K and $\tau_{imp}^{-1}=$ 200 meV in the superconducting state.
}
\end{figure}

All ground state DFT calculations are performed
at 400, 500 and 600 GPa, where metallic hydrogen 
is predicted to adopt the $\mathrm{I4_1/amd}$ 
crystal structure~\cite{PhysRevLett.112.165501,PhysRevLett.106.165302}.
We employ the QUANTUM-ESPRESSO package~\cite{0953-8984-21-39-395502}, 
using a plane-wave energy cutoff of 100 Ry and 
an ultrasoft pseudopotential\cite{PhysRevB.41.7892} 
with the Perdew-Wang parametrization of the 
local-density-approximation\cite{PhysRevB.45.13244} for the exchange and correlation potential. 
The wannierization process includes the 40 lowest-lying bands and is performed 
using the WANNIER90 package~\cite{Mostofi20142309}.
It allows us to interpolate the original $20\times20\times20$ \textbf{k}-space mesh
into a fine $60\times60\times60$ mesh for the calculation of the TDDFT 
dielectric function. In the latter crystal local field effects are taken into account by the use of 
two reciprocal lattice shells~\cite{PhysRevB.85.054305,PhysRevB.89.085102}.
The Eliashberg function $\alpha^2F(\omega)$ function needed to solve the
ME equations is calculated as described in Ref. \cite{PhysRevB.93.174308},
but with the use of the same exchange and correlation potential as
for the TDDFT calculation. $\alpha^2F(\omega)$ is calculated
at the harmonic level as anharmonicity barely affects it~\cite{PhysRevB.93.174308}. ME equations were solved
using a Matsubara energy cutoff of 6 times the highest phonon frequency, and the same as for computing the Padé approximant. We
solved the ME equations assuming different $\tau_{imp}^{-1}$ impurity scattering rates as the latter is introduced as a parameter in the equations.

\begin{figure}[t]
 \includegraphics[width=0.70\linewidth]{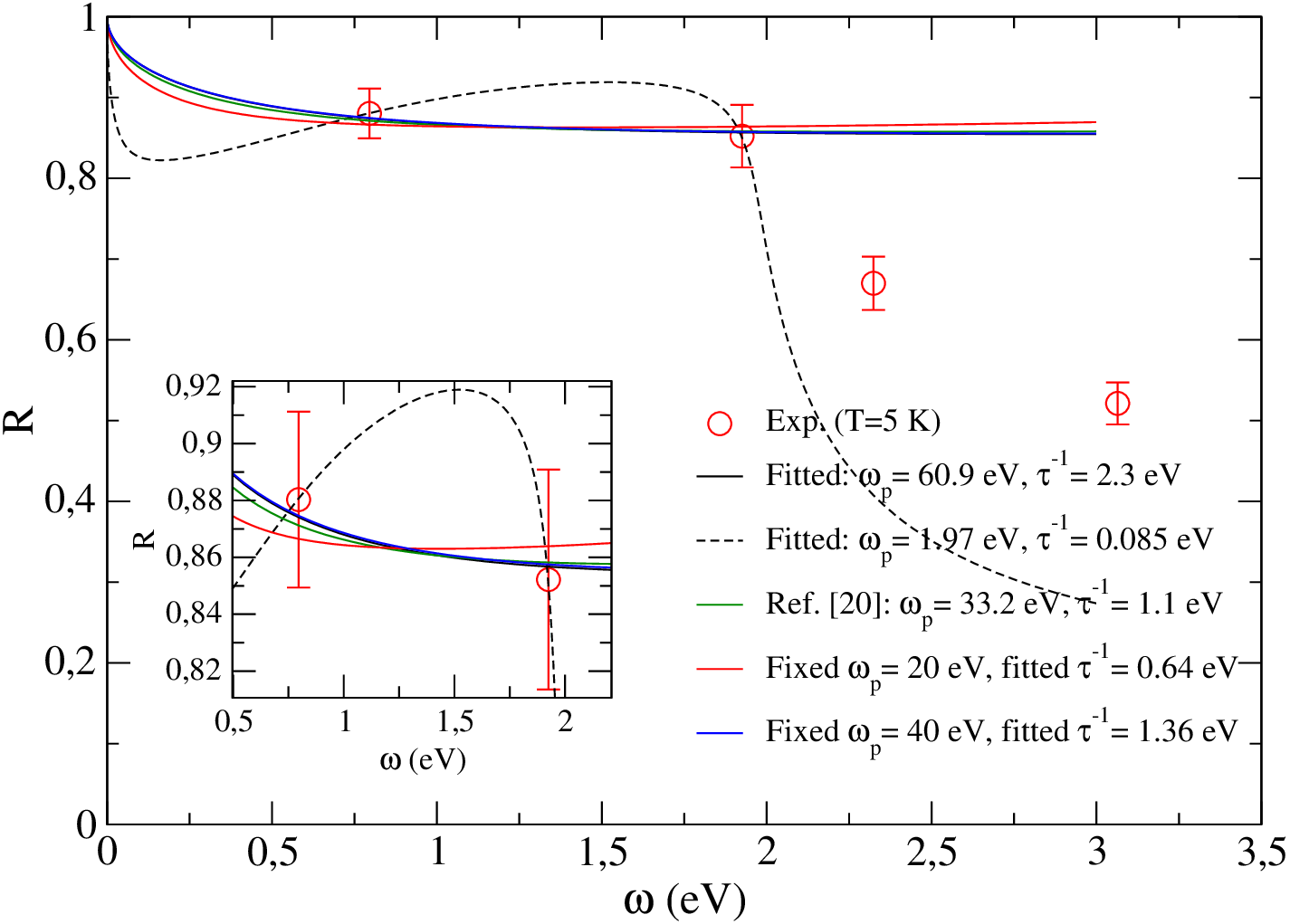}\hspace{0.1cm}
 \caption{\label{fitting} Experimental data at 5 K \cite{Diaseaal1579} fitted with the Drude model (only the two lowest energy data points are fitted). Solid and dashed black curves show our two different fitting results, while the green curve shows the fitting by
 Dias \textit{et al.}\cite{Dias2017}. Red and blue curves show the curves obtained by fitting only $\tau^{-1}$ for $\omega_p$ fixed at 20 and 40 eV respectively. 
}
\end{figure}

\section{Drude model}


In Ref. [\onlinecite{Diaseaal1579}] the authors fitted the experimental data with the reflectivity formula following the Drude model. According to this model,
\begin{equation}
\varepsilon(\omega)=1-  \frac{\omega_p^2 \tau}{\omega}\frac{1}{i+\omega\tau}.
\end{equation}

While often the plasma frequency $\omega_p$ and the mean scattering time $\tau$ are defined as fixed parameters, it can also be useful to generalize this formula by making them frequency dependent. This way, for a known
$\varepsilon(\omega)$, one can define $\omega_p(\omega)$ and $\tau(\omega)$ as

\begin{equation}
 \frac{1}{\tau(\omega)}=\frac{\omega \mathrm{Im}~\varepsilon(\omega)}{1-\mathrm{Re}~\varepsilon(\omega)}
\end{equation}

\begin{equation}
\omega_p^2(\omega)=\omega\tau(\omega)\left(\omega^2+\frac{1}{\tau(\omega)^2}\right)\mathrm{Im}~\varepsilon(\omega).
\end{equation}

We have calculated $\tau^{-1}(\omega)$ and $\omega_p(\omega)\equiv+\sqrt{|\omega_p^2(\omega)|}$ using $\varepsilon(\omega)$ values calculated for different impurity scattering rates and temperatures, as well as only considering TDDFT and therefore neglecting
phonon and impurity scattering. In Fig. \ref{tauwide} we can see that for frequencies larger than 15 eV, all $\omega_p$ curves converge in a plateau close to the theoretical $\omega_p=\sqrt{4\pi N}=$ 35.0 eV value, while
$\tau^{-1}$ yields around 7 eV, regardless of whether or not including phonon and impurity scattering. This shows electronic scattering is dominated by electronic band structure effects in this high frequency regime.
For frequencies lower than 5 eV but larger than 0.5 eV approximately, $\omega_p$ yields values around 21 eV, close to the $\omega_p=\sqrt{4\pi N_{intra}}=$ 22.6 eV value, which is expected as the interband transitions occur at higher energies and therefore their corresponding electronic density 
does not contribute to $\varepsilon$. In this low energy regime, $\tau^{-1}$ ranges 0.7-1.5 eV when phonon and impurity scattering is included in the calculation even if for the TDDFT only calculation yields negligible values. This shows phonon and impurities
clearly govern electronic scattering for the photon energies the experiment was performed at.  
In the 5-15 eV interband plasmon region, $\omega_p$ and $\tau^{-1}$ take unrealistic
(even negative for $\tau^{-1}$) values. This is because the Drude formula is not adequate for modeling the optical conductivity close to interband excitations. In order to take into account the interband transitions and their contribution to
$\varepsilon$, one should add a Lorentz Oscillator. The same holds for the very small energy region ($\omega<$ 0.5 eV), as the $\omega$ dependence of the electron-phonon and impurity scattering contribution to $\varepsilon$ is more complex
than the one modeled by the Drude formula.

\section{Fitting of experimental data}

\begin{table}[t]
\begin{tabular}{cccc}
\hline
\hline
Pressure (GPa)    &        400  &    500   &  600 \\
\hline
$\Delta_0$ (meV)  &       64.1  &   61.0   & 67.1 \\
$\omega_0$ (eV)   &        7.6  &    8.2   &  8.2 \\
$\omega_{inter}$ (eV)& 5.5  &    6.5   &  6.6 \\
$\omega_p$  (eV)   &       34.2  &   35.7   & 36.8 \\
\hline
\hline
\end{tabular}
\caption{Superconducting electronic bandgap ($\Delta_0$), interband absorption peak position ($\omega_0$), interband plasmon peak position ($\omega_{inter}$) and total plasma frequency ($\omega_p$) of $\mathrm{I4_1/amd}$ hydrogen at different pressures.
\label{changes}}
\end{table}

In Ref. [\onlinecite{Diaseaal1579}] the authors fitted four experimental data points for each temperature (5 and 83 K), where reflectivity was corrected for taking into account diamond absorption. However, later they claimed this correction procedure
may not be valid and thus only considered the lowest two energy raw (not corrected) data points\cite{Dias2017}, obtaining similar values for $\omega_p$ and $\tau^{-1}$ at 5 K. However, fitting a non-linear formula consisting of two parameters
to only two experimental points results in a non-unique solution for the fitting parameters. By fitting the first two experimental data points at 5 K to the reflectivity formula we have obtained at least two different results: $\omega_p=$ 
60.91 eV and 1.98 eV and with $\tau^{-1}=$ 2.21 eV and 0.085 eV respectively (see Fig. \ref{fitting}). None of the fitted $\omega_p$ values are reasonable, either for too high or too low. 
It is also important to notice we do not obtain the same fitting parameters as in Ref. [\onlinecite{Dias2017}] ($\omega_p=$ 33.2 eV and $\tau^{-1}=$ 1.1 eV). However,if we set $\omega_p=$ 33.2 eV and fit only $\tau^{-1}$ we obtain the same 1.1 eV as 
in Ref. [\onlinecite{Dias2017}].
We have seen that for $\omega_p$ values fixed within 20-40 eV (values close to what is expected for a nearly-free-electron-like metal at these densities) and fitted only $\tau^{-1}$ to the
experimental data, the latter only oscillates within 0.64-1.36 eV yielding a good fitting to the experiments within the error bars.   

\section{Pressure dependence}

Due to the doubts on the pressure calibration~\cite{Eremets2017,Loubeyre2017,Liu2017,Goncharov2017} in Ref. [\onlinecite{Diaseaal1579}] we have calculated the reflectivity of a $\mathrm{I4_1/amd}$ hydrogen/diamond interface at 400 and 600 GPa as well at 5 K and $\tau_{imp}^{-1}=$ 200 meV in the superconducting
state. As we can see in Fig. \ref{pressure}, differences are only quantitative, as the qualitative nature of the curves does not change. The minor quantitative changes of the main parameters are summarized in Table \ref{changes}.

\bibliography{bibliografia}
\bibliographystyle{apsrev4-1}